\documentclass[aps,pra,reprint,twocolumn,nofootinbib]{revtex4}

\usepackage{epsfig,amssymb,amsmath,physics,enumitem}
\usepackage{color,subfig,pdfpages}

\usepackage{braket}
\usepackage{xfrac}

\definecolor{myurlcolor}{rgb}{0,0,0.7}
\definecolor{myrefcolor}{rgb}{0.8,0,0}
\usepackage[unicode=true,pdfusetitle, bookmarks=true,bookmarksnumbered=false,bookmarksopen=false, breaklinks=false,pdfborder={0 0 0},backref=false,colorlinks=true, linkcolor=myrefcolor,citecolor=myurlcolor,urlcolor=myurlcolor]{hyperref}

\graphicspath{{./fig/}}

\def\equationautorefname~#1\null{%
  Eq.~(#1)\null
}

\newtheorem{theorem}{Theorem}
\def\theoremautorefname~#1\null{%
	Theorem~#1\null
}

\def\lemmaautorefname~#1\null{%
	Lemma~#1\null
}
\newtheorem{corollary}{Corollary}
\def\corollaryautorefname~#1\null{%
	Corollary~#1\null
}

\newcommand{\F}{\mathcal{F}}
\newcommand{\sys}{\mathrm{sys}}
\newcommand{\env}{\mathrm{env}}
\def\tr{\mbox{Tr}}

\renewcommand{\t}[1]{\textrm{#1}}
\newcommand{\thmref}[1]{\hyperref[#1]{Theorem~\ref{#1}}}
\newcommand{\lemmaref}[1]{\hyperref[#1]{Lemma~\ref{#1}}}
\newcommand{\figref}[1]{\hyperref[#1]{Fig.~\ref{#1}}}
\newcommand{\figaref}[1]{\hyperref[#1]{Fig.~\ref{#1}a}}
\newcommand{\figbref}[1]{\hyperref[#1]{Fig.~\ref{#1}b}}
\newcommand{\figcref}[1]{\hyperref[#1]{Fig.~\ref{#1}c}}
\renewcommand{\eqref}[1]{\hyperref[#1]{Eq.~(\ref{#1})}}
\newcommand{\eqsref}[2]{\hyperref[#1]{Eqs.~(\ref{#1})-(\ref{#2})}}
\newcommand{\appref}[1]{\hyperref[#1]{Appx.~\ref{#1}}}

\renewcommand{\var}{\phi}

\newcommand{\tphi}{{\tilde{\phi}}}

\newcommand{\n}{k}

\definecolor{teal}{rgb}{0.0, 0.5, 0.5}
\begin{document}
%\fbox{{\scriptsize Eprint: quant-ph/}}
%\fbox{{\scriptsize Preliminary draft from Lorenzo Maccone, \today}}
%\fbox{{\scriptsize Not for distribution.}}
%\fbox{{\scriptsize Submitted paper.}}

%\fbox{{\scriptsize Preliminary draft.}}
%Title of paper

\title{
	Number of bits returned by a quantum estimation
}
\author{
	Xi Lu$^1$, Wojciech Górecki$^2$,  Chiara Macchiavello$^{2,3}$, and Lorenzo Maccone$^{2,3}$}
\affiliation{
\vbox{
	$^1$School of Mathematical Science, Zhejiang University, Hangzhou, 310027, China
}\vbox{
	$^2$INFN Sez. Pavia, via Bassi 6, I-27100 Pavia, Italy
}\vbox{
	$^3$Dip. Fisica, University of Pavia, via Bassi 6, I-27100 Pavia, Italy
} \\ }

\begin{abstract}
    We give two upper bounds to the mutual information in arbitrary quantum estimation strategies. The first is based on some simple Fourier properties of the estimation apparatus. The second is derived using the first but, interestingly, depends only on the Fisher information of the parameter, so it is valid even beyond quantum estimation. We illustrate the usefulness of these bounds by characterizing the quantum phase estimation algorithm in the presence of noise. In addition, for the noiseless case, we extend the analysis beyond applying the bound and we discuss the optimal entangled and adaptive strategies, clarifying inaccuracies appearing on this topic in the literature.
\end{abstract}
%\pacs{03.65.Ta,06.30.Ft,03.65.Ud,03.67.-a}

%Time, measurement of, 06.30.Ft
%Spacetime topology of, 04.20.Gz
%Measurement theory (quantum mechanics), 03.65.Ta
%Quantum information, 03.67.-a
%Entanglement and quantum nonlocality, 
%Measurement theory (quantum mechanics), 03.65.Ta
%quantum mechanics, 03.65.Ta

\maketitle

% \section{Introduction}
Classical strategies, where the estimation is simply repeated $N$
times independently, can only achieve the standard quantum limit~(SQL)
in precision: the RMSE achieves the central limit scaling of
$N^{-1/2}$, and the Fisher information scales as $N$. Quantum
metrology~\cite{giovannetti2004quantum,giovannetti2006quantum,giovannetti2011advances,paris2009quantum,escher2011general,demkowicz2012elusive}
uses quantum properties, such as entanglement, to improve the
precision of parameter
estimation~\cite{caves1981quantum,mckenzie2002experimental,valencia2004distant,de2005quantum,budker2007optical,maze2008nanoscale}. In
the idealized noiseless case, many quantum strategies have been
proposed to achieve the Heisenberg scaling~(HS): the root mean square
error~(RMSE) scales as $N^{-1}$, or the Fisher information scales as
$N^2$. In the presence of noise the HS is rarely achieved, e.g. using
error correction \cite{dur2014improved}.  Although a scaling better
than SQL appears in some
scenarios~\cite{matsuzaki2011magnetic,chin2012quantum,jeske2014quantum,chaves2013noisy},
typically only the SQL can be
achieved~\cite{fujiwara2008fibre,escher2011general,demkowicz2012elusive},
and the quantum enhancement is only a constant factor.

Mutual information (MI) in quantum metrology is typically used in
connection to the
RMSE~\cite{hall2012heisenberg,hall2012does,sekatski2017improved}, but
it can also be used in a purely information-theoretic
way~\cite{hassani2017digital}. This allows one to easily account for
the prior information on the parameter, which is otherwise cumbersome
to deal with~\cite{tsang2012ziv,giovannetti2012sub}.  In this
scenario, the HS and the SQL refer to the cases when the mutual info
scales as $\log N$ and $\frac{1}{2}\log N$
respectively~\cite{hassani2017digital}. In the noiseless case, the HS
can be achieved using the quantum phase estimation~(QPE) algorithm,
and the SQL can be achieved by separable parallel strategies.

In this paper, we give two upper bounds (the ``Fourier bound'' and
the ``Fisher bound'') to the mutual information of quantum estimation,
for both periodic and infinite-range parameters.  The first bound
connects the mutual information to some Fourier properties of the
estimation procedure, and implies the second bound which is given in
terms of the Fisher information. As such, the second bound is still
valid in a more general setting, beyond quantum estimation.  We
illustrate some maximum-likelihood-based cases where the Fisher bound
is a good approximation to the mutual information.  Two applications
for these bounds are provided: a case study to show how the
bounds also work for noisy quantum estimation where the direct
calculation is practically impossible, and the analysis of the optimal general
separable and entangled strategies in the noiseless case (fixing some
inaccurate claims in the literature). We show some interesting
unknown features of the dephasing channel. In
\cite{Wojciech2024mutual} we found different bounds for the MI, derived
from purely mathematical considerations. In contrast, knowledge of the
estimation procedure is required here for the derivation of the Fourier
bound  (and the Fisher bound is, in turn, derived from it). We compare
all these bounds showing no one is tighter than the others in all
situations.

Finally, for the noiseless case,  we go beyond the application of the bound, performing the detailed analysis of optimal entangled and adaptive strategies.

\section{Fourier bound}

Consider a quantum channel that encodes
a parameter $\phi$ onto a pure normalized state $\ket{\psi_\phi}$ (e.g.~a
unitary encoding), followed by any positive operator-valued measurement~(POVM). The following holds:

\begin{theorem}\label{thm:fourier} ``Fourier bound''.
  Given a parameterized quantum state $\phi\mapsto\ket{\psi_\phi}$,
  with $\phi$ having period $L$ and prior density $p(\phi)$, and any POVM that returns some classical information $m$, the
  mutual information between $\phi$ and $m$ satisfies,
	\begin{equation}
		I(m:\phi) \leq -\sum_{k=-\infty}^{\infty} \hat{f}_k \log \hat{f}_k - \log L + H(\phi),
		\label{eq:mi_periodic}
	\end{equation}
	where
	$
		H(\phi) := -\int  \dd\phi\:p(\phi) \log p(\phi)
	$ is the entropy of $\phi$,
	$
		\hat{f}_k = \frac{1}{L} \ip{\hat{\psi}_k}
	$,
	\begin{equation}
		\ket{\hat{\psi}_k} := \int \dd\phi\: q(\phi) e^{-i \frac{2\pi k}{L} \phi} \ket{\psi_\phi},
		\label{eq:def_hat_psi}
	\end{equation}
	for any complex-valued function $q(\phi)$ such that $q(\phi)q^*(\phi) = p(\phi)$.
\end{theorem}
The RHS of (\ref{eq:mi_periodic}) is independent of the measurement that returns $m$ because the inequality uses the Holevo bound~\cite{Holevo1982} which is valid for any measurement.
Throughout the paper, all integrations are over one period of the periodic variable.
Note that the RHS of \autoref{eq:mi_periodic} depends on the choice of $q(\phi)$, which counterbalances the arbitrary global phase of $\ket{\psi_\phi}$.
The bound is not tight, as the Holevo bound is used in the proof, which is not tight in general.

\textit{Proof.}
Construct another parameterized quantum state,
\begin{equation}
	\theta\mapsto\ket{\Psi_\theta} := \int \dd\phi' \: q(\phi'+\theta) \ket{\psi_{\phi'+\theta}} \ket{\phi'},
\end{equation}
where $\ket{\phi}$ lives in an additional $L^2(\mathbb{R})$ rigged Hilbert space with
$
\ip{\phi_2}{\phi_1}=\delta(\phi_2-\phi_1)
% (\int \bra{\tphi}a^*(\tphi)\dd\tphi)(\int b(\phi)\dd\phi\ket{\phi})=\int a^*(\phi)b(\phi)\dd\phi
$.
We can measure the first register the same way as the original state, and the second register on the computational basis, to obtain a joint measurement result, say $(m,\phi')$.
The probabilities of the additional $\theta$-parameterized state (with subscripts $i$) and the original $\phi$-parameterized state (with subscripts $o$) are related via,
\begin{equation}
	p_i(m,\phi|\theta) = p(\phi+\theta) p_o(m|\phi+\theta),
\end{equation}
where $p$ without subscripts is for the prior distribution.
Let the parameter $\theta$ in the new state be distributed uniformly in $[0,L]$, then the joint distribution of $m$ and $\phi$ in the new state is,
\begin{equation}
\begin{aligned}
	p_i(m,\phi)
	= &
	\frac{1}{L} \int \dd\theta\: p(\phi+\theta) p_o(m|\phi+\theta) 
	\\ = &
	\frac{1}{L} \int \dd\phi\: p(\phi) p_o(m|\phi)
	=
	\frac{1}{L} p_o(m).
\end{aligned}
\end{equation}

The mutual information between measurement result $(m,\phi)$ and the parameter $\theta$ is,
\begin{equation}
\begin{aligned}
	&
	I_i(m,\phi:\theta)
	 = 
	\sum_m \int \dd\theta \dd\phi\: \frac{1}{L} p_i(m,\phi|\theta)
	\log \frac{p_i(m,\phi|\theta)}{p_i(m,\phi)}
	\\ = &
	\frac{1}{L} \sum_m \int \dd\theta \int \dd\phi\: \:p(\phi+\theta) p_o(m|\phi+\theta)
	\\ & \cdot
	\log \frac{p(\phi+\theta) p_o(m|\phi+\theta)}{
		\frac{1}{L} p_o(m)
	}
	\\ = &
	\frac{1}{L} \sum_m \int \dd\theta \int \dd\phi\: p(\phi) p_o(m|\phi)
	\log \frac{L p(\phi) p_o(m|\phi)}{
		p_o(m)
	}
	\\ = &
	\sum_m \int \dd\phi\: p(\phi) p_o(m|\phi)
	\log \frac{L p(\phi) p_o(m|\phi)}{
		p_o(m)
	}
	\\ = &
	I_o(m:\phi) + \int \dd\phi\: p(\phi) \log [L p(\phi)],
\end{aligned}
\label{eq:rel_two_circ}
\end{equation}
where the last equality uses $\sum_m p_o(m|\phi) = 1$. We
can use the Holevo bound~\cite{Holevo1982} which says
that the mutual information $I(m:\phi)$ between the parameter and
any measurement is upper bounded by the Holevo-$\chi$ quantity
\begin{equation}
	\chi = 
	S \left(\int \dd\phi\: \rho_\phi p(\phi)  \right)
	-
	\int \dd\phi\: S(\rho_\phi) p(\phi),
	\label{eq:chi_qpe}
\end{equation}
where $S(\rho)=-\tr(\rho\log\rho)$ is the von Neumann entropy,
$\rho_\phi$ is the density matrix of the channel output state, and the
integral is over a period when $\phi$ is periodic, or over the real
line when $\phi$ has infinite range. Applying Holevo's bound to the
new state we get $ I_i(m,\phi:\theta) \leq S(\rho) $, where
$ \rho = \frac{1}{L} \int \dyad{\Psi_\theta} \dd\theta $, and the
second term in Holevo's bound vanishes since pure states
$\dyad{\Psi_\theta}$ have zero von Neumann entropy.

Since $\ket{\Psi_\theta}$ is $L$-periodic in $\theta$, we can expand it in a Fourier series
$
	\ket{\Psi_\theta} = \sum_{k} \ket{\hat{\Psi}_k} e^{i\frac{2\pi k}{L}\theta}
$, where
\begin{equation}
\begin{aligned}
	\ket{\hat{\Psi}_k}
	= &
	\frac{1}{L} \int q(\phi+\theta) e^{-i\frac{2\pi k}{L}\theta} \ket{\psi_{\phi+\theta}} \ket{\phi} \dd\theta\dd\phi
	\\ = &
	\frac{1}{L} \int q(\theta') e^{-i\frac{2\pi k}{L}(\theta'-\phi)} \ket{\psi_{\theta'}} \ket{\phi} \dd\theta'\dd\phi
	\\ = &
	\ket{\hat{\psi}_k} \left(
		\frac{1}{L} \int e^{i\frac{2\pi k}{L}\phi} \ket{\phi} \dd\phi
	\right).
\end{aligned}
\end{equation}

Moreover,
\begin{equation}
\begin{aligned}
	&
	\ip{\hat{\Psi}_j}{\hat{\Psi}_k}
	\\ = &
	\frac{1}{L^2}
	\ip{\hat{\psi}_j}{\hat{\psi}_k}
	\int \dd\phi'\dd\phi e^{i\frac{2\pi}{L}(k\phi-j\phi')} \ip{\phi'}{\phi}
	\\ = &
	\frac{1}{L^2}
	\ip{\hat{\psi}_j}{\hat{\psi}_k}
	\int \dd\phi e^{i\frac{2\pi}{L}(k-j)\phi}
	\\ = &
	\hat{f}_k \delta_{jk},
\end{aligned}
\end{equation}
where $\delta_{jk}$ is the Kronecker delta.

Finally, from
\begin{equation}
\begin{aligned}
	\rho
	= &
	\frac{1}{L} \int \dd\theta \sum_{j,k=-\infty}^{\infty} e^{i\frac{2\pi}{L}(j-k)\theta} \dyad{\hat{\Psi}_j}{\hat{\Psi}_k}
	\\ = &
	\sum_{k=-\infty}^{\infty} \dyad{\hat{\Psi}_k},
\end{aligned}
\label{eq:rho_spectrum}
\end{equation}
we know that $\rho$ has eigenvalues $\{\hat{f}_k\}$ with eigenstates $\{\ket{\hat{\Psi}_k}\}$, thus
\begin{equation}
	% I(m,\phi:\theta)
	% \leq
	S(\rho)
	=
	-\sum_{k=-\infty}^{\infty} \hat{f}_k \log \hat{f}_k.
	\label{eq:rho_eigen}
\end{equation}

Combining \autoref{eq:rel_two_circ} and \autoref{eq:rho_eigen}, we have the desired result.
\hfill $\Box$

\begin{corollary}
\label{cor:entrunc}
\autoref{thm:fourier} applied to the standard problem of the phase estimation with uniform prior and covariant measurement, results in the well-known entropic uncertainty relations~\cite{bialynicki1975uncertainty,hall1993phase}
\begin{equation}
\label{eq:entrounc}
   H(\tphi-\phi)+H(|c_k|^2)\geq 0,
 \end{equation}
 where $H(\tphi-\phi)$ and $H(|c_k|^2)$ are the entropies of phase and number measurements
(and where the parametrization chosen here removes the constant sometimes
   found on the RHS).
\end{corollary}

Indeed, consider the family of states
\begin{equation}
\label{eq:entstate}
 \ket{\psi_\var}={\textstyle\sum_{k=0}^\infty} c_k e^{ik 2\pi\var}\ket{k},
\end{equation}
where $\phi\in[0,1)$ and the (uniform) prior is $p(\phi)=1$ (see \autoref{sec:uniatry} for the broader context). Then $\forall_k\ket{\hat \psi_k}=\ket{k}$, so the bound gives
\begin{equation}
\label{eq:simbo}
I(m:\phi) \leq -\sum_{k=0}^{\infty} |c_k|^2 \log |c_k|^2.
\end{equation}

Next, note that the mutual information may be written as
$I(m:\phi)=H(\phi)-H(\phi|m)$, where $H(\phi|m)$ is the conditional entropy. Finally, for the covariant measurement, the POVM is given by 
\begin{equation}
\label{eq:covmea}
\{M_{\tphi}\}_{\tphi}=\{\ket{\tphi}\bra{\tphi}\}_{\tphi},\quad \t{with}\ \ket{\tphi}={\textstyle\sum_{k=0}^\infty} e^{i k 2\pi\tphi}\ket{k},
\end{equation}
(where the measurement outcome is labeled directly but the value of corresponding estimator $\tilde\var$ instead of $m$), we have 
$H(\phi|m)=H(\phi|\tphi)=H(\tphi-\phi)$. As here $H(\phi)=0$, we have $I(\tphi:\phi)=-H(\tphi-\phi)$, so the bound for MI is exactly equivalent to \eqref{eq:entrounc}.

The results in this section can be extended to non-periodic case, see \appref{app:non_periodic}.

\section{Fisher bound}
We now use the Fourier bound above to derive the Fisher one. Since the Fisher bound is independent of the quantum encoding, it is valid in a more general setting as follows:

\begin{theorem}\label{thm:fisher} ``Fisher bound''.
    Consider a general parameter estimation problem that estimates a parameter $\phi$ (with prior density $p$) from a conditional probability density $p(m|\phi)$.
	If $\phi$ has period $L$, then
	\begin{equation}
	\begin{aligned}
		I(m:\phi) \leq & \frac{1}{2} \log(1 + \frac{eL^2}{8 \pi}\int \left[
			\frac{\dot{p}(\phi)^2}{p(\phi)} + p(\phi) \F(\phi)
			\right] \dd\phi)
		\\ & - \log L + H(\phi),
	\end{aligned}
	\label{eq:mi_fisher_periodic}
	\end{equation}
 
    where $\F(\phi) = \sum_m {\dot{p}(m|\phi)^2}/{p(m|\phi)}$ is
    the Fisher information, with
    $\dot{p}(m|\phi)=\partial p(m|\phi)/\partial\phi$.
\end{theorem}

The proof relies on \autoref{thm:fourier} and as a consequence, the Fisher bound is not tight, either.

\textit{Proof.}
Consider the quantum parameter estimation problem $\phi\mapsto\sum_m\sqrt{p(m|\phi)}\ket{m}$, with the prior density $p(\phi)$.
If we measure on the computational basis, we obtain the same probability distribution as the original problem, thus they share the same mutual information, and any bound on the mutual information of the quantum estimation problems also works on the original problem.
Using the notations from \autoref{thm:fourier}, in which we choose $q(\phi)=\sqrt{p(\phi)}$,
\begin{equation}
\begin{aligned}
	\hat{f}_k
	= &
    \frac{1}{L}
	\left\|
		\sum_{m} \int \sqrt{p(\phi)p(m|\phi)} e^{-i \frac{2\pi k}{L} \phi} \ket{m} \dd\phi
	\right\|^2
	\\ = &
    \frac{1}{L}
	\sum_m \left|
		\int \sqrt{p(\phi)p(m|\phi)} e^{-i \frac{2\pi k}{L} \phi} \dd\phi
	\right|^2,
\end{aligned}
\end{equation}
where $\|\ket{\psi}\|^2 := \ip{\psi}$.

Define
$
	\sigma^2
	:=
	\sum k^2 \hat{f}_k
$, then
\begin{equation}
\begin{aligned}
	\sigma^2
	= &
    \frac{1}{L}
	\sum_{m,k} \left|
		\int \sqrt{p(\phi)p(m|\phi)} ke^{-i \frac{2\pi k}{L} \phi} \dd\phi
	\right|^2
	\\ = &
	\frac{L^3}{4\pi^2}
	\sum_{m,k} \left|
        \frac{1}{L}
		\int \left[ \frac{\dd}{\dd\phi}\sqrt{p(\phi)p(m|\phi)} \right] e^{-i \frac{2\pi k}{L} \phi} \dd\phi
	\right|^2
	\\ = &
	\frac{L^2}{4\pi^2}
	\sum_m \int \left|
		\frac{\dd}{\dd\phi}\sqrt{p(\phi)p(m|\phi)}
	\right|^2 \dd\phi
	\\ = &
	\frac{L^2}{16\pi^2}
	\int \sum_m \left[
		\frac{\dot{p}(\phi)^2}{p(\phi)} p(m|\phi) + 2\dot{p}(\phi) \dot{p}(m|\phi)
    \right. \\ & \left.
        + p(\phi) \frac{\dot{p}(m|\phi)^2}{p(m|\phi)}
	\right] \dd\phi
	\\ = &
	\frac{L^2}{16\pi^2}
	\int \left[
		\frac{\dot{p}(\phi)^2}{p(\phi)} + p(\phi) \F(\phi)
	\right] \dd\phi,
\end{aligned}
\label{eq:sigma2}
\end{equation}
in which the third equality uses the identity $
    \sum_k |\hat{F}_k|^2 = \frac{1}{L} \int |F(\phi)|^2 \dd \phi
$ for any period-$L$ function $F(\phi)$ and its Fourier coefficients $\hat{F}_k=\frac{1}{L}\int F(\phi) e^{-i \frac{2\pi k}{L} \phi} \dd\phi$, and the last equality uses $\sum_m p(m|\phi) = 1$ and $\sum_m \dot{p}(m|\phi) = 0$.

\begin{figure}
	\centering
	\includegraphics[width=.45\textwidth]{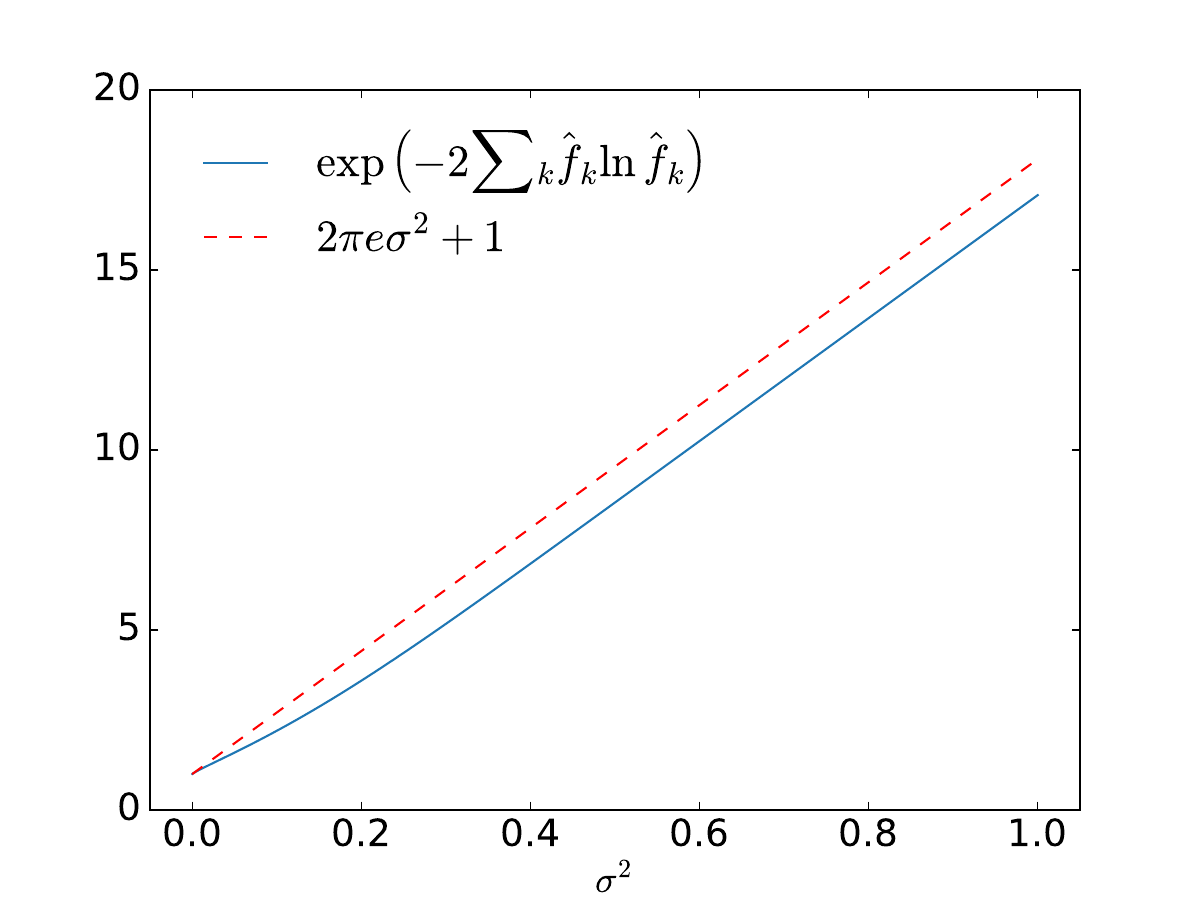}
	\caption{Numerical results (solid line) to prove the upper bound (dash line) in \autoref{eq:chi_sigma}. }
	\label{fig:numerical_sigma}
\end{figure}

By Lagrange's multiplier method, the optimal $\hat{f}_k$ that maximizes $-\sum_k \hat{f}_k \log \hat{f}_k$ with constraints $\sum_k \hat{f}_k = 1$ and $\sum_k k^2\hat{f}_k = \sigma^2$ is the Gauss-like sequence,
\begin{equation}
	\hat{f}_k = \frac{1}{\sqrt{2\pi}c} e^{-\frac{k^2}{2b^2}},
	\label{eq:opt_f_periodic}
\end{equation}
for some $b,c>0$.
One should determine $b$ and $c$ by the two constraints.
In \autoref{fig:numerical_sigma}, we numerically calculate the optimal $-\sum_k \hat{f}_k \log \hat{f}_k$ using \autoref{eq:opt_f_periodic}.
These numerical results suggest,
\begin{equation}
	-\sum_k \hat{f}_k \log \hat{f}_k \leq \frac{1}{2}\log(1+2\pi e\sigma^2).
	\label{eq:chi_sigma}
\end{equation}
although this inequality is not strictly proven analytically.

Combining with \autoref{eq:sigma2} and \autoref{thm:fourier}, we obtain \autoref{eq:mi_fisher_periodic}.
\hfill $\Box$

\begin{figure}
	\centering
	\includegraphics[width=.45\textwidth]{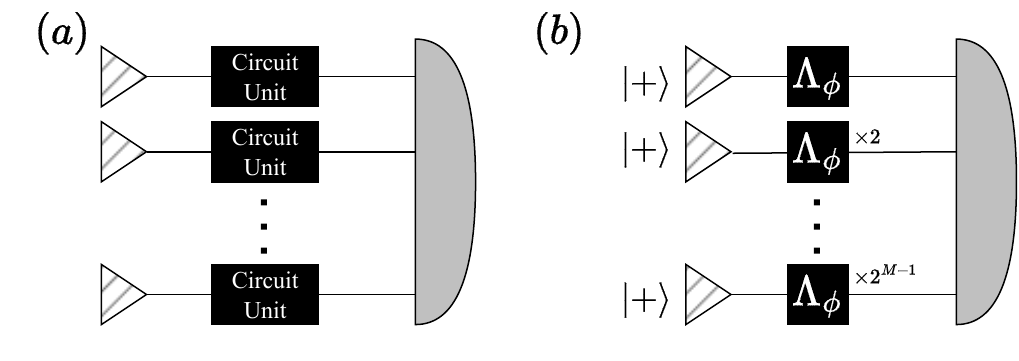}
	\caption{(a) Repeated circuits; (b) Quantum phase estimation
        (QPE), where the inverse quantum Fourier transform is
        included in the measurement and is not shown explicitly.}
	\label{fig:special_pattern}
\end{figure}

To show how closely the bound \autoref{eq:mi_fisher_periodic} can be
achieved, consider the case of a quantum estimation where $N$
circuit units are used independently on separately prepared probes,
see \autoref{fig:special_pattern}(a).  For simplicity, we assume
the uniform prior $p(\phi)=\frac{1}{L}$ and constant Fisher
information $\F$ in the circuit unit.  The total Fisher information is
$N\F$.  The bound is therefore simplified to,
\begin{equation}
	I(m:\phi) \leq \frac{1}{2} \log(1 + \frac{e L^2}{8\pi} N\F).
\end{equation}
If different $\phi$'s lead to different states in the circuit unit, then the maximum likelihood estimator is asymptotically efficient, and we can obtain a lower bound to the mutual information~\cite{brunel1998mutual},
\begin{equation}
\begin{aligned}
	I(m:\phi)
	\gtrsim &
	H(\phi)
	-
	\int\dd\phi p(\phi) \frac{1}{2} \log\frac{2\pi e}{N\F}
	\\ = &
	\log L - \frac{1}{2} \log\frac{2\pi e}{N\F}
	=
	\frac{1}{2} \log\frac{L^2N\F}{2\pi e}.
\end{aligned}
\label{eq:mi_lower}
\end{equation}

In the large $N$ limit, the difference between the upper and lower bounds is approximately 
$\log\frac{e}{2}\approx 0.44~\mathrm{bits}$.

 We now compare \autoref{thm:fisher} with the Fisher
information based bound given in \cite{Wojciech2024mutual}.  On one
hand, when $L=1$, $p(\phi)=1$ and $\F(\phi)=\F$ is constant over
$\phi$, \autoref{thm:fisher} gives,
\begin{equation}
    I(m:\phi)
    \leq
    \frac{1}{2} \log\left(
        1 + \frac{e}{8\pi} \F
    \right)
    \leq
    \log\left(
        1 + \sqrt{\frac{e}{8\pi}} \sqrt{\F}
    \right),
\end{equation}
which is slightly tighter than,
\begin{equation}
    I(m:\phi)
    \leq
    \log\left(
        1 + \frac{1}{2} \sqrt{\F}
    \right),\label{other}
\end{equation}
from \cite{Wojciech2024mutual}. On the other hand, when $p(\phi)$ has
discontinuities, there are delta function terms appearing in
$\dot{p}(\phi)$, which make the right hand side of
\autoref{eq:mi_fisher_periodic} diverge, while the bound (\ref{other})
can still converge. Namely, the two types of bounds,
(\ref{eq:mi_fisher_periodic}) and (\ref{other}), complement each
other in different situations.

Note also that, as discussed in \cite{Wojciech2024mutual}, Efroimovich’s inequality \cite{efroimovich1980information,aras2019family,lee2022new} cannot be applied for $L$-periodic problem in its basic form. See \cite{lee2022new} for its generalization for log-concave priors, which requires much more complicated mathematical formalism.

\section{Noisy phase estimation}

Now we analyze the case of phase estimation in the presence of noise.   Obtaining the performance of quantum metrology in the presence of noise is extremely cumbersome already for the RMSE  \cite{fujiwara2008fibre,escher2011general,demkowicz2012elusive,matsuzaki2011magnetic,chin2012quantum,jeske2014quantum,chaves2013noisy},  but the bounds presented above allow us to give sophisticated bounds also for the MI.

\begin{figure}
    \centering
    \includegraphics[width=0.45\textwidth]{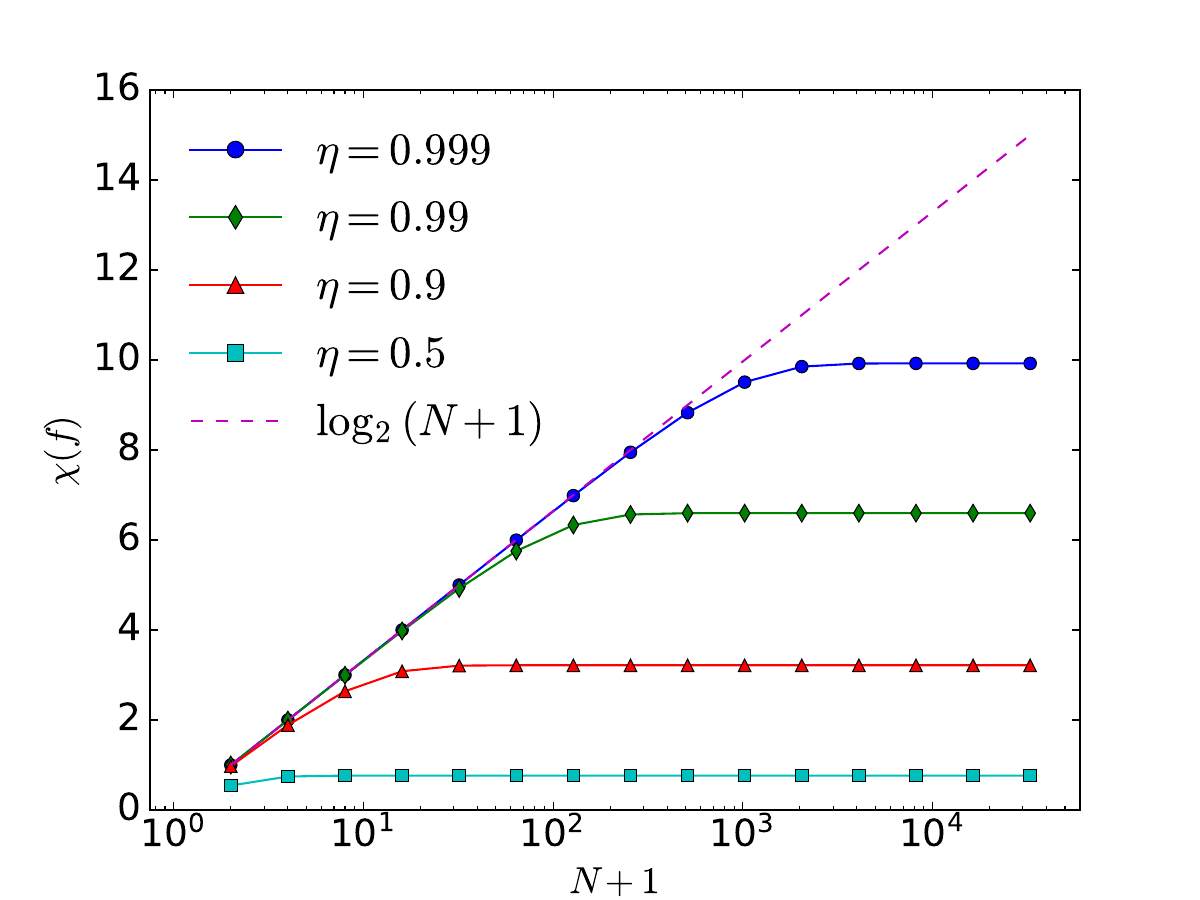}
    \caption{ The mutual information upper bound for noisy QPE. }
    \label{fig:chi_qpe}
\end{figure}

%\section{Example of noisy quantum phase estimation}

%We now use the bounds derived in the previous section to  analyze the highly nontrivial case of noisy quantum estimation.
  
As an example, we consider the QPE algorithm with $M$-qubits,
shown in \autoref{fig:special_pattern}(b).  Although the theorems
work only for pure states, they can be used in the presence of noise
by purifying the output state. Start with the dephasing
channel
$\Lambda_\phi(\rho)=U_\phi(\frac{1+\eta}2\rho+\frac{1+\eta}2\sigma_z\rho\sigma_z)U^\dag_\phi$
with $U_\phi=\dyad{0}+e^{i 2\pi\phi}\dyad{1}$, $\eta$ the noise
parameter.  The density
matrix before the measurement is
\begin{equation}
	\rho = \otimes_{j=0}^{M-1} \frac{1}{2} \begin{bmatrix}
		1 & (\eta e^{-i 2\pi\phi})^{2^{j}} \\
		(\eta e^{i 2\pi\phi})^{2^{j}} & 1
	\end{bmatrix}.
\end{equation}

\begin{figure}
    \centering
    \includegraphics[width=0.45\textwidth]{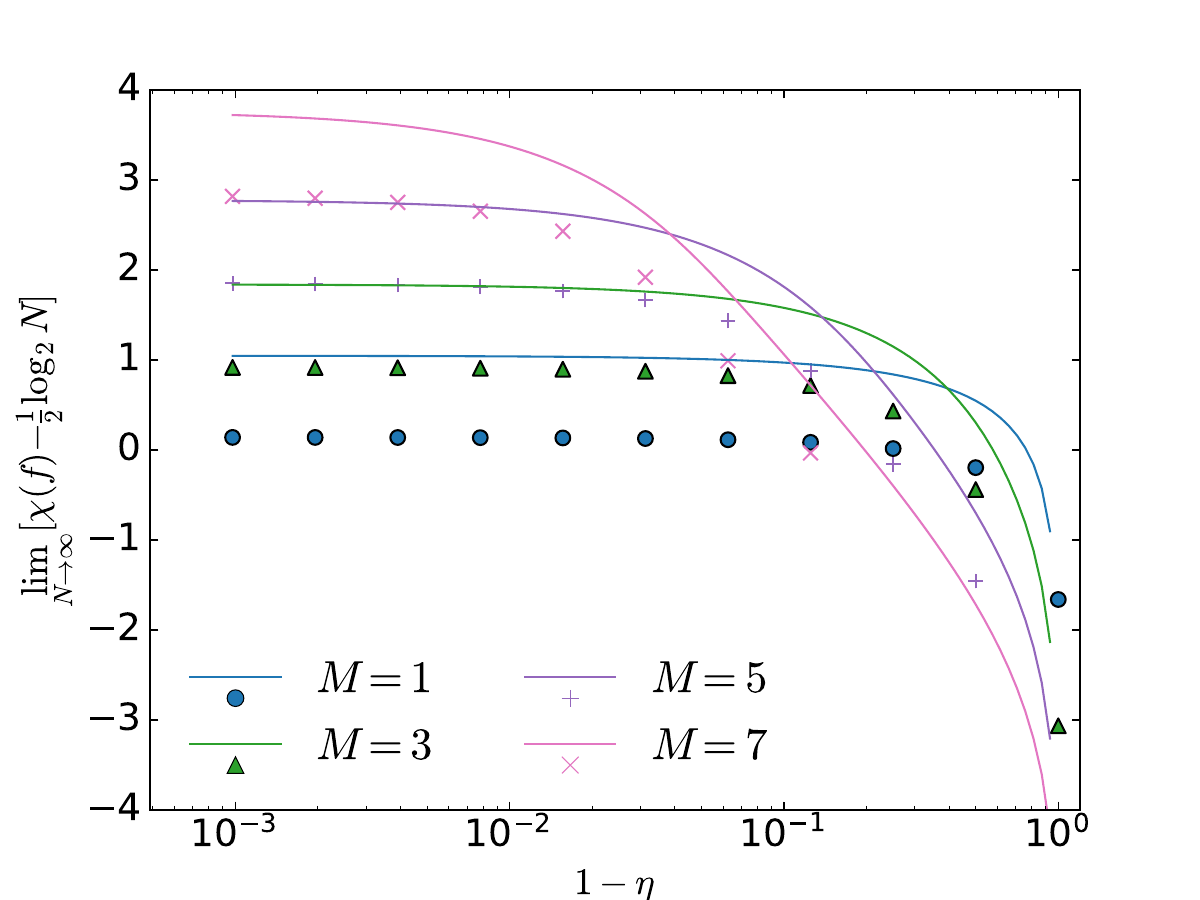}
    \caption{Quantum enhancement, namely the second term in the
          RHS of \eqref{eq:asympt_mi}, for the dephasing QPE as a
          function of the noise, are shown as the continuous lines. The markers of the same colors and detached from the lines show that mutual information calculated by numerical experiments.} 
    \label{fig:repeated_qpe}
\end{figure}

It can be purified as
\begin{equation}
\begin{aligned}
	\ket{\psi_\phi} = & \otimes_{j=0}^{M-1} \frac{1}{\sqrt{2}} \left[
		\ket{0}_{\sys}\ket{0}_{\env}
		+
		(\eta e^{i 2\pi\phi})^{2^{j}} \ket{1}_{\sys}\ket{0}_{\env}
	\right. \\ & \left.
		+
		\sqrt{1 - \eta^{2^{j+1}}} \ket{1}_{\sys}\ket{1}_{\env}
	\right],
\end{aligned}
\label{eq:purification_qpe}
\end{equation}
with \textit{env} a purification space.
This is a unitary encoding $\ket{\psi_\phi}=e^{i\phi H}\ket{\psi_0}$ for some Hermitian operator $H$.
By \autoref{thm:fourier}, $\{\hat{f}_k\}$ are the Fourier coefficients of the function $f(\phi)=\ip{\psi_0}{\psi_\phi}$ such that $f(\phi)=\sum_k\hat{f}_k e^{i 2\pi k\phi}$, and
$
    I(m:\phi) \leq \chi(f) := -\sum_{k=-\infty}^{\infty} \hat{f}_k \log \hat{f}_k
$.
Here,
\begin{equation}
\begin{aligned}
	& f(\phi) = \prod_{j=0}^{M-1} \left(1-\frac{\eta^{2^{j+1}}}{2} + \frac{(\eta e^{i 2\pi\phi})^{2^{j}}}{2}\right),
    \\ \Rightarrow	& \chi(f) =
	\sum_{j=0}^{M-1} H_{\mathrm{bin}} \left(
		\frac{\eta^{2^{j+1}}}{2}
    \right),
  \end{aligned}
\end{equation}
where $H_{\mathrm{bin}}(x)=-x\log x-(1-x)\log(1-x)$ is the binary
entropy function. In \autoref{fig:chi_qpe} we show $\chi$ as a
  function of $N$ for different noise levels $\eta$. In the low $N$
  and low noise regime, each new qubit provides approximately one
  extra bit of mutual information as expected. But for larger $N$, the
  gain of extra qubits tends to zero and $\chi$ converges to a
  constant value in the asymptotic limit even in the presence of
  slight noise. This implies that, in the presence of noise, the QPE
  should use smaller QFT units, and repeat the procedure several
times, instead of using a single large QFT circuit. Namely the circuit
in \autoref{fig:special_pattern}(a), in which the circuit unit is
\autoref{fig:special_pattern}(b).  In this case, the total number of
calls to the phase gate is $N=R(2^M-1)$ where $M$ is the number of
qubits in the circuit unit, and $R$ the number of repetitions. In the
large $N$ limit, \autoref{thm:fisher} gives,
\begin{equation}
	I(m:\phi) \lesssim \frac{1}{2} \log N + \frac{1}{2} \log \frac{e \F}{8\pi(2^M-1)},
	\label{eq:asympt_mi}
\end{equation}
where $\F=(2\pi)^2 \sum_{j=0}^{M-1} 4^j \eta^{2^{j}}$ is its constant
quantum Fisher information.  This is an SQL asymptotic scaling,
with a constant factor which encodes the quantum enhancement and
which is plotted in \figref{fig:repeated_qpe}. As expected, larger QFT units perform better in the small-noise
region, but also are more affected in the large-noise region.  As
the noise level increases, the optimal strategy for the dephasing
changes from the full-size QPE to the classical separate strategy
($M=1$).
In comparison, we numerically calculated the mutual information of \label{fig:special_pattern}(a) with 10 repetitions of the QPE unit and the final estimation is obtained by Bayesian estimation, shown as the markers in \figref{fig:repeated_qpe}.
The calculation uses Monte Carlo importance sampling on $m$ in the definition of $I(m:\phi)$, since there is an intractable amount of measurement results in this setting and it is computationally impossible to calculate it explicitly.
The comparisons between each pair of upper bound and experimental results show that the bound \autoref{eq:asympt_mi} can characterize correctly how the mutual information behaves in the noisy setting, and the easy-to-compute upper bound can be a nice indicator on the choice of optimal noisy estimation scheme, compared to the hard-to-compute mutual information.

Qualitatively similar results hold also for the amplitude
  damping and erasure noise, see \appref{app:qpe_other_channels}.

\section{Discussion of noiseless estimation beyond the bounds}
\label{sec:uniatry}
%{\it Heisenberg scaling in phase estimation.---}

\begin{figure}[t!]
\includegraphics[width=0.45\textwidth]{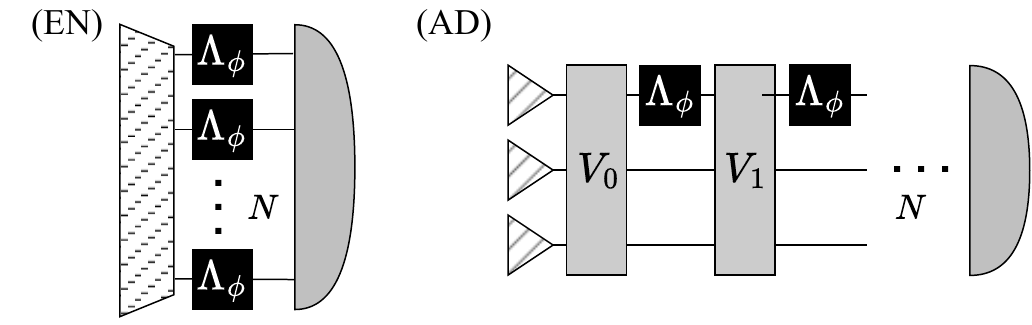}
\caption{Different possible quantum estimation schemes, using $N$ elementary gates jointly:  starting from an entangled input state
  (shaded box) in EN or in an adaptive way in AD ($V_i$ representing joint unitaries). EN may be seen as a special case of AD}\label{fig:all}
\end{figure}
\begin{figure}[t!]
\includegraphics[width=0.45\textwidth]{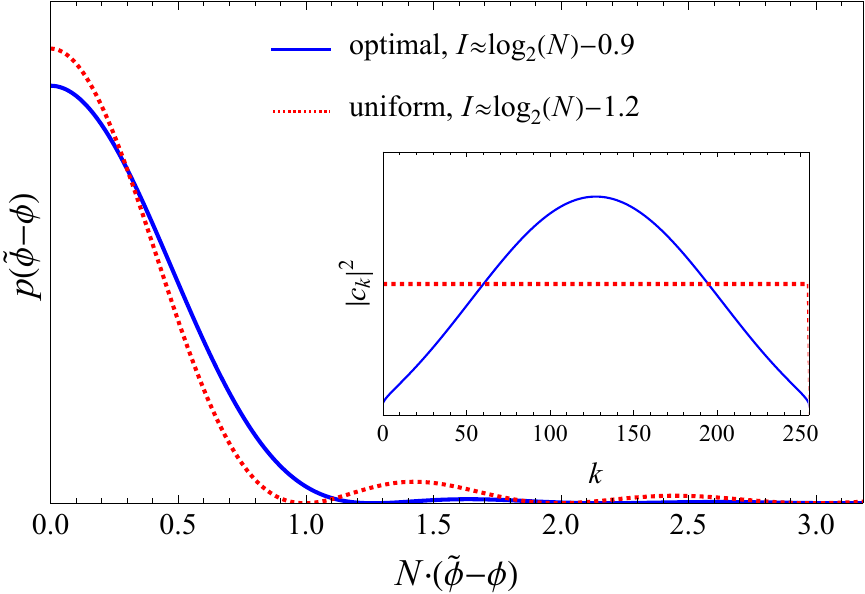}
\caption{Posterior distributions $p(\tphi-\phi)$ for the EN
 protocol with different input states, the optimal one and the one with uniform weights ($N=255$ here). Inset: square modulus of the amplitudes of the above
  states. %For sufficiently large $N$ $p(\phi-\var)$ does not change the shape significantly, is only rescaled by factor $N$. .
}
\label{fig:opt}
\end{figure}

In this section, we discuss an optimal joint use of all $N$ gates in the noiseless case with uniform prior. Start with the case where they are used in parallel, acting on a single entangled input state, see EN in \autoref{fig:all}. Without loss of the generality, we may restrict to fully symmetric input states. Then, the output state is fully characterized by the number $k$ of bits in the $\ket{1}$ position, via
\begin{equation}
\label{eq:entstateN}
 \ket{\psi_\var}={\textstyle\sum_{k=0}^N} c_k e^{ik 2\pi\var}\ket{k}.
\end{equation}
(Compared to \eqref{eq:entstate}, here the maximal value of index index $k$ is set to be $N$). As observed in corollary \ref{cor:entrunc}, \autoref{thm:fourier} applied to this problem results in bounding the MI by the entropy of the $|c_k|^2$ coefficients
\begin{equation}
\label{eq:simboN}
I(m:\phi) \leq -\sum_{k=0}^{N} |c_k|^2 \log |c_k|^2.
\end{equation}
The input state maximizing the bound is the one with uniform weights $\frac{1}{\sqrt{N+1}}\sum_{n=0}^N\ket{n}$. For fixed covariant measurement:
\begin{equation}
\label{eq:covmeaN}
\{M_{\tphi}\}_{\tphi}=\{\ket{\tphi}\bra{\tphi}\}_{\tphi},\quad \t{with}\ \ket{\tphi}={\textstyle\sum_{k=0}^N} e^{i k 2\pi\tphi}\ket{k},
\end{equation}
MI is shown to be equal to $-H(\tphi-\phi)$. The analytical minimization of $H(\tphi-\phi)$ is nontrivial and no
general formula is known for arbitrary $N$. However, such minimization
is simple to do by numerical means, see Fig.~\ref{fig:opt}. Interestingly, the optimal state turns out to be significantly different than the one maximizing $H(|c_k|^2)$, which follows from the fact, that the entropic uncertainty relation (\ref{eq:entrounc}) is not tight in general.

Now, is the measurement (\ref{eq:covmeaN}) indeed optimal for the maximizing MI in the EN scheme? Can these results be overcome by applying a more advantage adaptive scheme (AD in \autoref{fig:all}), where the gates are used sequentially acting on the state entangled with ancilla, with arbitrary unitary controls acting between?

%Let us start with the entangled scheme (EN). Restricting to the fully symmetric space, we may characterize the output state as:
%\begin{equation}
% \ket{\psi_\var}={\textstyle\sum_{\n=0}^N} c_\n e^{i\n2\pi\var}\ket{n},
%\end{equation}
%where $\ket{n}$ is fully symmetric normalized states of $N$ qubits, where $n$ of them occupy state $\ket{1}$, and $N-n$ the state $\ket{0}$. The measurement which is commonly believed to be the optimal is a covariant one:
%\begin{equation}
%\label{eq:covmea}
%\{M_\phi\}_\phi=\{\ket{\phi}\bra{\phi}\}_{\phi},\quad \t{with}\ \ket{\phi}={\textstyle\sum_{\n=0}^N} e^{i\n2\pi\phi}\ket{\n},
%\end{equation}
While the measurement \eqref{eq:covmeaN} is commonly believed to be optimal, its optimality has been not proven formally yet. This measurement is known to be optimal for minimizing standard Bayesian cost~\cite{Holevo1982}, for which it is also proven, that its performance cannot be overcome by any AD strategy ~\cite{chiribella2008memory}. However, in the case of MI the optimality of single-seed covariant measurements has been proven only for irreducible group representations \cite{chiribella2006optimal,davies1978information} or for a single use of the $U_\var$ gate~\cite{sasaki1999accesible} (more details in \appref{app:davies}).
%~\footnote{See also   \cite{lu2012hierarchy,micadei2015coherent,zhou2020saturating} for a   discussion on the necessity of global measurements in the case of   entangled states.}

%At last, one more comment may be than about SEQ-L without local communication (so standard QPEA cannot be applied). In analyzing the Bayesian cost minimization problem, it was observed that for optimal performance it is useful to repeat different sequences of usage $k_i=2^i$ different times (keeping the total number of gates used $N$). Especially, for the number of repetition decreasing linearly with $i$, the HS was formally shown~\cite{higgins2009demonstrating}. Therefore for this protocol the relation~\eqref{eq:varentr} HS for MI as well.

Here we show, that also in the MI case a full optimization over the whole
protocol AD resorts to an optimization over the input state in the
entangled protocol EN, using the measurement of
\eqref{eq:covmeaN}. Indeed, even if we are not able to prove the
general optimality of this measurement for any input state, we can
show that it is optimal, if also the state is optimal.

\begin{theorem}
  Consider the most general estimation protocol for phase estimation with a uniform prior $p(\var)=1$,
  (AD) of Fig.~\ref{fig:all}. Then, the optimal performance is the
  same as the best one obtainable for EN using the covariant
  measurement of \eqref{eq:covmeaN}:
  \begin{equation}
\max_{(AD)} I=\max_{(EN)} I=\max_{\ket{\psi}}I[\ket{\psi},\{\ket{\tilde\phi}\bra{\tilde\phi}\}_{\tilde\phi}],
\end{equation}
where in the RHS $I[\ket{\psi},\{\ket{\tilde\phi}\bra{\tilde\phi}\}_{\tilde\phi}]$ indicates the MI from the POVM (\ref{eq:covmeaN}) on
the state $\ket{\psi}$.
\end{theorem}
This simplifies the whole optimization of finding the best strategy to
an optimization only over the input state's $N+1$ real coefficients.
  (The question of whether this measurement is
optimal also for a broader class of input states remains
open. Numerics support this conjecture, but we have no general analytical
proof.)

\textit{Proof.}  The proof is inspired by reasoning
from~\cite{Gorecki2020pi}, applied to Bayesian estimation
there. First, from the convexity of MI, the optimal input state is a
pure state and the optimal measurement is the rank-one measurement
(see App. \ref{app:fullopt} for a detailed justification). 

Next, note that the output state
for any adaptive strategy may be written as % (here we use the notation $\prod_{i=0}^NA_i=A_NA_{N-1}...A_0$ for ordered operators product):
\begin{equation}
\label{eq:ada}
\begin{split}
\ket{\psi_\var}&=\prod_{i=1}^N\nolimits V_i(\sum_{b_i\in\{0,1\}}\nolimits e^{ib_i2\pi\var}\ket{b_i}\bra{b_i}\otimes\openone)V_0\ket{\psi}\\
&=\sum_{\{b_1,b_2,...b_N\}}\nolimits\prod_{i=1}^N\nolimits V_i(e^{ib_i2\pi\var}\ket{b_i}\bra{b_i}\otimes\openone)V_0\ket{\psi}\\
&=\sum_{\n=0}^N e^{i\n2\pi\var}\underbrace{\left[\sum_{\{b_i\}:\sum b_i=\n}\prod_{i=1}^N V_i(\ket{b_i}\bra{b_i}\otimes\openone)V_0\ket{\psi}\right]}_{=:c_\n\ket{g_\n}}
\end{split}
\end{equation}
where $\openone$ acts on all ancillas, the $\ket{g_\n}$ are
normalized, but not necessarily orthogonal (namely,
$\int_{-\infty}^{+\infty} \dd\var\: e^{-ip2\pi\var}\ket{\psi_\var}$ is non
zero only for $p\in[0,1,...,N]$), and $\ket{\psi}$ is the global initial
state.

Then,
since $I(x,\var)=H(\var)-H(\var|x)$, for fixed prior the maximization of MI is
equivalent to minimization of the
conditional entropy $H(\var|x)$, which may be seen as the average of
entropies of the posterior distribution. Therefore, it may be bounded
from below by the minimal possible entropy of a posterior distribution
obtained for a single element, rank-one measurement
$\ket{\chi}\bra{\chi}$:
\begin{equation}
\label{eq:minapo}
  H(x|\var)\geq\min_{\ket{\psi},\{V_i\}_i,\ket{\chi}}
  -\int d\var\: p(\var|\chi)
  {\log} p(\var|\chi),
\end{equation}
where $p(\var|\chi)=\mathcal N_\chi |\braket{\chi|\psi_\var}|^2$, with
$\mathcal N_\chi$ the normalization. This allows us to write
$p(\var|\chi)=\mathcal N_\chi
|\braket{\chi|\psi_\var}|^2=|\sum_{\n=0}^N e^{i\n2\pi\var}d_n|^2$,
with $d_\n:=\sqrt{\mathcal N_\chi}c_n\braket{\chi|g_n}$. Then the
RHS of \eqref{eq:minapo} may be further bounded by:
\begin{equation}
    \geq\min_{d_n}-\int d\var |\sum_{\n=0}^N\nolimits e^{i\n2\pi\var}d_\n|^2{\log} |\sum_{\n=0}^N\nolimits e^{i\n2\pi\var}d_\n|^2
\end{equation}
with $\sum_{\n=0}^N |d_\n|^2=1$.  Moreover, for the optimal
performance, all $d_\n$ may be chosen real.  Finally, note that, after
finding the optimal $d_\n$, the inequality (\ref{eq:minapo}) may be
saturated by taking an entangled input state
$\ket{\psi}=\sum_\n d_\n\ket{n}$ and the covariant measurement 
(\ref{eq:covmeaN}), without the need for any adaptivity.
\hfill $\square$

\section{Conclusions}

In conclusion, we have given two types of upper bounds to the
  mutual information: (i) the Fourier bounds given by
  \autoref{eq:mi_periodic} for periodic parameters (phase);
  % , and by \autoref{eq:mi_infinite} for infinite-range parameters with finite prior support;
  and (ii) the Fisher bound of
  \autoref{eq:mi_fisher_periodic}, which is written only in terms of
  the Fisher information of the parameter, rather than on the details
  of the estimation, so it applies even beyond quantum estimation, where it has been derived above. As
  shown in \autoref{eq:mi_lower}, the bound in terms of Fisher
  information for periodic parameters are asymptotically a good
  approximation to the mutual information (namely, it is almost
  attainable) for the maximum likelihood estimator. These bounds are
  useful to characterize the noisy QPE algorithm.

   X.L. acknowledges the National Natural Science Foundation of China under Grant Nos.62272406, 61932018 and Zhejiang University for funding, and the University of Pavia for
   hospitality. L.M. C.M. W.G.~acknowledge financial support from the
   U.S. DoE, National Quantum Information Science Research Centers,
   Superconducting Quantum Materials and Systems Center (SQMS) under
   contract number DE-AC02-07CH11359, C.M. acknowledges support from the EU H2020 QuantERA project QuICHE and from the PNRR MUR Project PE0000023-NQSTI. L.M. acknowledges support from the
   PNRR MUR Project CN0000013-ICSC, and from PRIN2022 CUP 2022RATBS4.

\bibliographystyle{unsrt}
\bibliography{ref.bib}

% % Supplemental material
% \clearpage
% \onecolumngrid
% \begin{center}
% 	\textbf{\large Supplemental Material for ``How many bits does your noisy quantum estimation return?"}
% 	\vspace{0.5cm}
% 	\\Xi Lu$^1$, Lorenzo Maccone$^2${\red,  [...]}
% 	\\\textit{
% 		$^1$School of Mathematical Science, Zhejiang University, Hangzhou, 310027, China
% 		\\$^2$Dip.~Fisica and INFN Sez.~Pavia, University~of Pavia, via Bassi 6, I-27100 Pavia, Italy
% 	}
% \end{center}

\appendix
\section{Non-periodic estimation}\label{app:non_periodic}

\autoref{thm:fourier} was derived for periodic
  parameters (phase estimation). It can be extended to cases with
  unbounded range but finite support of the prior:

\begin{theorem}\label{thm:holevo_infinite}
  Given a parameterized quantum state $\phi\mapsto\ket{\psi_\phi}$
  with prior density $p(\phi)$ having finite support, then the
  mutual information between the measurement result $m$ and the
  parameter $\phi$ satisfies,
	\begin{equation}
		I(m:\phi) \leq -\int \hat{f}(k)\log\hat{f}(k) \dd k + H(\phi),
		\label{eq:mi_infinite}
	\end{equation}
	where $\hat{f}(k) = \ip{\hat{\psi}(k)}$, with
        $\ket{\hat{\psi}(k)} := \int q(\phi) e^{-i 2\pi k \phi}
        \ket{\psi_\phi} \dd\phi$.\end{theorem}

\textit{Proof.}
Pick large enough $L$ such that $p$ vanishes outside an interval of length $L$.
We can treat $p$ as the periodic case, and obtain from \autoref{thm:fourier},
\begin{equation}
	I(m:\phi)
	\leq
	-\frac{1}{L}
	\sum_{k\in\frac{1}{L}\mathbb{Z}} \hat{f}(k) \log \hat{f}(k)
	+ H(\phi),
\end{equation}
in which we use $\sum_{k\in\frac{1}{L}\mathbb{Z}} \frac{1}{L} \hat{f}(k) = 1$.
Taking $L\to\infty$, we get the desired result.
\hfill $\Box$

A similar argument also holds for the Fisher bound
  (\autoref{eq:mi_fisher_periodic}), but one obtains a bound that, in this
  case, is strictly looser than Efroimovich’s inequality \cite{efroimovich1980information,aras2019family,lee2022new}.

\section{Quantum phase estimation with other noisy channels}\label{app:qpe_other_channels}

For the amplitude damping noise,
\begin{equation}
	\Lambda_\phi(\rho) := \begin{bmatrix}
		\rho_{00} + (1-\eta) \rho_{11}           & \sqrt{\eta} e^{-i 2\pi\phi} \rho_{01} \\
		\sqrt{\eta} e^{i 2\pi\phi} \rho_{10}     & \eta \rho_{11}
	\end{bmatrix}.
\end{equation}

The density matrix before the measurement is,
\begin{equation}
	\rho = \otimes_{j=0}^{M-1} \frac{1}{2} \begin{bmatrix}
		2-\eta^{2^{j}}                         & (\sqrt{\eta} e^{-i 2\pi\phi})^{2^{j}} \\
		(\sqrt{\eta} e^{i 2\pi\phi})^{2^{j}}   & \eta^{2^{j}}
	\end{bmatrix}.
\end{equation}

We use the purification,
\begin{equation}
\begin{aligned}
	\ket{\psi_\phi} = & \otimes_{j=0}^{M-1} \frac{1}{\sqrt{2}} \left[
		\sqrt{2-\eta^{2^{j}}} \ket{0}_{\sys}\ket{0}_{\env}
    \right. \\ & \left.
    	+
    	\frac{(\sqrt{\eta}e^{i2\pi\phi})^{2^{j}}}{\sqrt{2-\eta^{2^{j}}}} \ket{1}_{\sys}\ket{0}_{\env}
    \right. \\ & \left.
    	+
    	\sqrt{\eta^{2^{j}} \frac{1-\eta^{2^{j}}}{2-\eta^{2^{j}}}} \ket{0}_{\sys}\ket{1}_{\env}
	\right].
\end{aligned}
\end{equation}

Then,
\begin{equation}
	f(\phi) = \prod_{j=0}^{M-1} \left(
        1 - \frac{\eta^{2^{j}}}{4-2\eta^{2^j}}
    	+
    	\frac{(\eta e^{i 2\pi\phi})^{2^{j}}}{4-2\eta^{2^j}}
     \right),
\end{equation}
and
\begin{equation}
	\chi(f)
	=
	\sum_{j=0}^{M-1} H_{\mathrm{bin}} \left(
		\frac{\eta^{2^{j}}}{4-2\eta^{2^j}}
	\right).
\end{equation}

For the erasure noise,
\begin{equation}
	\Lambda_\phi(\rho) := \begin{bmatrix}
		\eta \rho_{00}                  & \eta \rho_{01} e^{-i 2\pi\phi}    & 0 \\
		\eta \rho_{10} e^{i 2\pi\phi}   & \eta \rho_{11}                    & 0 \\
		0                               & 0                                 & 1-\eta
	\end{bmatrix},
\end{equation}
where a third dimension is added indicating qubit loss.

The density matrix before the measurement is,
\begin{equation}
	\rho = \otimes_{j=0}^{M-1} \frac{1}{2} \begin{bmatrix}
		\eta^{2^{j}}                      & (\eta e^{-i 2\pi\phi})^{2^{j}}      & 0 \\
		(\eta e^{-i 2\pi\phi})^{2^{j}}      & \eta^{2^{j}}                      & 0 \\
		0                                   & 0                                   & 2(1 - \eta^{2^{j}})
	\end{bmatrix}.
\end{equation}

We use the purification,
\begin{equation}
\begin{aligned}
	\ket{\psi_\phi} = & \otimes_{j=0}^{M-1} \left[
		\eta^{2^{j-1}} \frac{\ket{0}_{\sys}+e^{i 2^{j+1}\pi\phi}\ket{1}_{\sys}}{\sqrt{2}} \ket{0}_{\env}
    \right. \\ & \left.
    	+
    	\sqrt{1 - \eta^{2^{j}}} \ket{2}_{\sys} \ket{1}_{\env}
	\right].
\end{aligned}
\end{equation}

Then,
\begin{equation}
	f(\phi) = \prod_{j=0}^{M-1} \left(
        1 - \frac{\eta^{2^{j}}}{2}
    	+
    	\frac{(\eta e^{i 2\pi\phi})^{2^{j}}}{2}
     \right),
\end{equation}
and
\begin{equation}
	\chi(f)
	=
	\sum_{j=0}^{M-1} H_{\mathrm{bin}} \left(
		\frac{\eta^{2^{j}}}{2}
	\right).
\end{equation}

\section{About the non-applicability of Davies theorem}
\label{app:davies}

For the general problem of maximizing the mutual information in a group estimation with a covariant prior, the optimal POVM is always guaranteed to be the one of the form: \cite{davies1978information,chiribella2006optimal}
:
\begin{equation}
\{M_{i,g}\}_{i,g}=\{\tfrac{1}{|G|}U_g^\dagger A_iU_g\}_{i,g},
\end{equation}
where $A_i$ (called seed) is a one-rank operator. In \cite{chiribella2006optimal} the minimal number of seeds that guarantees the optimal performance has been derived, depending on the group representation properties. Especially, the sufficiency of using a single seed $A_i$ is guaranteed only in the case where the group representation is irreducible (Davies theorem \cite{davies1978information}), which is not the case discussed in this paper. Indeed, to see the essence of the problem, consider the question of whether we can find a better measurement than
\begin{equation}
\label{eq:uno}
    \{U^\dagger_{\tphi} \ket{\chi}\bra{\chi} U_{\tphi}\}_{\tphi},\quad \t{with}\quad \ket{\chi}=\sum_{n=0}^n \ket{n},
\end{equation}
for which $p(\tphi|\phi)=\tr(\rho_{\phi-\tphi}\ket{\chi}\bra{\chi})$.
Consider the case in which one has two seeds:
\begin{equation}
\ket{\chi_1}=\sum_{n=0}^N a_n \ket{n},\quad \ket{\chi_2}=\sum_{n=0}^N b_n \ket{n},
\end{equation}
where $\forall_n |a_n|^2+|b_n|^2=1$.
Then the set of operators:
\begin{equation}
\label{eq:due}
    \{U_{\tphi}^\dagger \ket{\chi_i}\bra{\chi_i} U_{\tphi}\}_{i,\tphi}
\end{equation}
is a proper POVM, with probability distribution given as $q(i,\tphi|\phi)=\tr(\rho_{\phi-\tphi}\ket{\chi_i}\bra{\chi_i})$. Note, that the label $i$ gives no information on $\phi$, as $\lambda_i=\int d\tphi\: q(i,\tphi|\phi)$ is $\phi$-independent.

We may even construct the realization of the POVM where $i$ is drawn randomly before measuring $\tphi$. To do that, we need to consider an ancillary system, such that the output state is given by
\begin{equation}
\ket{\psi_\phi}=\sum e^{i2\pi n\phi}c_n\ket{n}\otimes \ket{1}.
\end{equation}
Then we apply the unitary operation
\begin{multline}
U=\sum_{n=0}^N\ket{n}\bra{n}\otimes\\ (a_n^*\ket{1}\bra{1}+b_n^*\ket{2}\bra{1}-b_n\ket{2}\bra{1}+a_n\ket{2}\bra{2}),
\end{multline}
getting as a result:
\begin{equation}
\begin{split}
\ket{\psi'_\phi}=&\sum e^{i2\pi n\phi}a^*_nc_n\ket{n}\otimes \ket{1}\\
+&\sum e^{i2\pi n\phi}b^*_nc_n\ket{n}\otimes \ket{2}.
\end{split}
\end{equation}
Next we perform a projective measurement on the ancillary system on a basis $\{\ket{1},\ket{2}\}$. Then, the probabilities are given by $\lambda_1=\sum |a_n^*c_n|^2$, $\lambda_2=\sum |b_n^*c_n|^2$, which are clearly independent on $\phi$. Finally, after performing the POVM \eqref{eq:uno}  on the remaining post-measurement state, we have globally realized \eqref{eq:due}.
Define
\begin{equation}
q(i,\tphi,\phi)=\lambda_i q(\tphi,\phi|i).
\end{equation}
Even if for each $i$, $\{U_{\tphi}^\dagger \ket{\chi_i}\bra{\chi_i} U_{\tphi}\}_{\tphi}$ by itself does not form a POVM, the probabilities $q(\tphi,\phi|1),q(\tphi,\phi|2)$ are well defined, and they should be understood as the probabilities obtained in the post-selection process.

To compact the notation, from now on by $I\left[p(\tphi,\phi)\right]$ we indicate the mutual information between the variables $\tphi,\phi$, for the joint probability $p(\tphi,\phi)$.
From the convexity of MI, 
\begin{multline}
\label{eq:fromconv}
I\left[\lambda_1q(\tphi,\phi|1)+\lambda_2q(\tphi,\phi|2)\right]\\
\leq\lambda_1 I(q(\tphi,\phi|1))+\lambda_2 I(q(\tphi,\phi|2)),
\end{multline}
where the LHS corresponds to the POVM $\{U^\dagger_{\tphi}(\ket{\chi_1}\bra{\chi_1}+\ket{\chi_2}\bra{\chi_2})U_{\tphi}\}_{\tphi}$
i.e. the one where the experimentalist has lost knowledge about $i$, while the RHS corresponds to $\{U_{\tphi}^\dagger \ket{\chi_i}\bra{\chi_i} U_{\tphi}\}_{i,\tphi}$.

One can easily prove that, without knowledge of $i$, it would indeed perform worse than \eqref{eq:uno}:
\begin{equation}
\label{eq:always}
\forall_{\ket{\chi_1},\ket{\chi_2}} I\left[\lambda_1q(\tphi,\phi|1)+\lambda_2q(\tphi,\phi|2)\right]\leq I\left[p(\tphi,\phi)\right],
\end{equation}
which is in favor of \eqref{eq:uno}
But it is easy to find an example for which:
\begin{equation}
\exists_{\ket{\chi_1}} I[p(\tphi,\phi)]\leq I[q(\tphi,\phi|1)]
\end{equation}
which is in favor of \eqref{eq:due}.

However, the choice of $\ket{\chi_1}$ imposes conditions on the $\ket{\chi_2}$. As a result, it is not clear if there exists a pair $\ket{\chi_1}$, $\ket{\chi_2}$ (satisfying $\forall_n |a_n|^2+|b_n|^2=1$) for which:
\begin{equation}
\label{eq:wonder}
(\exists_{\ket{\chi_1},\ket{\chi_2}}?) I(p(\tphi,\phi))<\lambda_1 I(q(\tphi,\phi|1))+\lambda_2 I(q(\tphi,\phi|2)).
\end{equation}
Numerical calculations suggest that there is no such a pair, and \eqref{eq:uno} is indeed optimal. So far, for all examples checked numerically, the inequality \eqref{eq:wonder} holds in the opposite way.

Once again, an analogous problem does not appear in the case of Bayesian cost, as then the analog of \eqref{eq:fromconv} is an equality, so \eqref{eq:always} would imply the falsity of \eqref{eq:wonder}.

\subsection*{Regarding the Lemma for real representation}

In~\cite{hassani2017digital} the Authors proposed using Lemma 2 from~\cite{sasaki1999accesible} to extend Davies theorem \cite{davies1978information} to the problem of phase estimation (which is an reducible representation). However, while it works for the single gate case~\cite{sasaki1999accesible}, it cannot be easily applied to a larger number of gates, as we show below.

Lemma 2 from \cite{sasaki1999accesible} is as follows.
For input state $\ket{\psi}$ and group representation $U_g$, assume that there exists a basis $\mathcal H=\t{span}_{\mathbb C}\{\ket{v_i}\}_i$, such that both $\ket{\psi}$ and $U_g$ has only real coefficients in this basis. Then if $U_g$ acting on $\t{span}_{\mathbb R}\{\ket{v_i}\}_i$ is a \textbf{real irreducible representation}, one may apply Davies' theorem.

For further discussion, it is worth noticing that our problem of measuring $\phi$ for state $\ket{\psi_\phi}=\sum_{\n=0}^N c_\n e^{i\n2\pi\phi}\ket{k}$ may be seen as measuring the rotation of the spin-$N/2$ particle, so fRom now we will use the notation connected with angular momentum.

To investigate the consequences of Lemma 2, we choose the basis in which the evolution is generated by $J_y$ (which is purely imagined), so $e^{i2\pi\phi J_y}$ is purely real. 

For spin-$1/2$, if we start from the real state $\ket{\psi}=\cos(\theta)\ket{-\tfrac{1}{2}}+\sin(\theta)\ket{+\tfrac{1}{2}}$, it evolves as $\ket{\psi_\phi}=\cos(\theta+\pi\phi)\ket{-\tfrac{1}{2}}+\sin(\theta+\pi\phi)\ket{+\tfrac{1}{2}}$, so U(1) indeed acts irreducible on $\t{span}_{\mathbb R}\{\ket{-\sfrac{1}{2}},\ket{+\sfrac{1}{2}}\}$, as noticed in~\cite{sasaki1999accesible}. However, it stops working for higher spins.

For example, for spin-$1$, the $J_y$ matrix has the following form:
\begin{equation}
   J_y=\frac{1}{\sqrt{2}}
   \begin{bmatrix}
0 & -i & 0\\
i & 0 & -i \\
0 & i & 0
\end{bmatrix},
\end{equation}
with eigenvectors:
\begin{equation}
  \ket{+1}_y=   \frac{1}{2}\begin{bmatrix}
-1 \\
-i\sqrt{2}\\
1 
\end{bmatrix},
  \ket{0}_y=   \frac{1}{\sqrt{2}}\begin{bmatrix}
1 \\
0\\
1 
\end{bmatrix},
  \ket{-1}_y=   \frac{1}{2}\begin{bmatrix}
-1 \\
i\sqrt{2}\\
1 
\end{bmatrix}.
\end{equation}
Therefore, we may distinguish two real subspaces, on which $e^{i2\pi \phi J_y}$ acts irreducibly:
\begin{equation}
\begin{split}
    V_1&=\t{span}_{\mathbb R}\{\ket{0}_y\}=\t{span}_{\mathbb R}\{\tfrac{1}{\sqrt{2}}(\ket{-1}_z+\ket{+1}_z)\},\\
    V_2&=\t{span}_{\mathbb R}\{\tfrac{1}{\sqrt{2}}(\ket{-1}_y+\ket{+1}_y),\tfrac{i}{\sqrt{2}}(\ket{-1}_y-\ket{+1}_y)\}\\
    &=\t{span}_{\mathbb R}\{\tfrac{1}{\sqrt{2}}(\ket{-1}_z-\ket{+1}_z),\ket{0}_z\},
    \end{split}
\end{equation}
while the whole representation is clearly reducible. Especially, if, following~\cite{hassani2017digital}, we consider the product state of two spin-$1/2$ particles, oriented perpendicular to the axis of rotation:
\begin{multline}
\ket{\phi}^{\otimes 2}=\left[\tfrac{1}{\sqrt{2}}(\sin(\pi\phi)\ket{+\tfrac{1}{2}}_z+\cos(\pi\phi)\ket{-\tfrac{1}{2}}_z)\right]^{\otimes 2}\\
=
\sin^2(\pi\phi)\ket{+1}_z+\sqrt{2}\sin(\pi\phi)\cos(\pi\phi)\ket{0}_z\\
+\cos^2(\pi\phi)\ket{-1}_z,
\end{multline}
we see, that it belongs solely neither to $V_1$ nor $V_2$, so Lemma 2 from~\cite{sasaki1999accesible} cannot be applied here.

\section{Optimality of rank-one measurement and pure input state}
\label{app:fullopt}

First, we argue that, for maximizing MI, we may always restrict to rank-one measurements.
Indeed, assume by contradiction that some of the $\{M_k\}_k$ is not rank-one. Then for each $M_k$, we consider its eigendecomposition $M_k=\sum_i M_k^i \ket{m_k^i}\bra{m_k^i}$ and construct a rank-one POVM of the form $\{M_k^i \ket{m_k^i}\bra{m_k^i}\}_{i,k}$. It is clear, that the new POVM has at least the same mutual information as the original one, as any statistic obtained from $\{M_k\}_k$ may be also obtained from $\{M_k^i \ket{m_k^i}\bra{m_k^i}\}_{i,k}$ by neglecting information related to the  index $i$. Note that, in general, $\{M_k^i \ket{m_k^i}\bra{m_k^i}\}_{i,k}$ does not need to be a projective measurement (as $M_k^i$ may be smaller than one), but this not affects the following reasoning.

Next, the fact that for any fixed POVM $\{M\}_k$ the minimum is obtained for pure states comes directly from the convexity of Shannon entropy:
\begin{multline}
\label{eq:shannon}
      H(\lambda p_1+(1-\lambda)p_2)\geq \lambda H(p_1)+ (1-\lambda) H(p_2)\\
      \geq \min\{H(p_1),H(p_2))\}
\end{multline}
 and the linearity of the Born rule, e.i. for $\rho=\sum_i x_i\ket{\psi_i}\bra{\psi_i}$:
\begin{equation}
      p(x|\phi)=\tr(\rho_\phi M_x)=\sum_i x_i\underbrace{\tr(\ket{\psi_{i}}\bra{\psi_i}_\phi M_x)}_{p_i(x|\phi)}.
\end{equation}

\end{document}